\documentclass[12pt]{article}
\usepackage{graphicx}
\usepackage{tabularx}
\usepackage{epstopdf}
\usepackage{epsfig}
\usepackage{amsmath}
\usepackage{amsfonts}
\usepackage{amssymb}
\usepackage{colordvi}
\usepackage{multirow}
\usepackage{subfigure}
\usepackage{rotating}
\usepackage{array}
\def\lsim{\mathrel{\raise.3ex\hbox{$<$\kern-.75em\lower 1ex\hbox{$\sim$}}}}
\def\gsim{\mathrel{\raise.3ex\hbox{$>$\kern-.75em\lower 1ex\hbox{$\sim$}}}}

\def\be{\begin{equation}}
\def\ee{\end{equation}}
\def\bea{\begin{eqnarray}}
\def\eea{\end{eqnarray}}

\begin{document}

\title{COHERENT constraints on \\nonstandard neutrino interactions }

\author{Jiajun Liao and Danny Marfatia\\
\small\it Department of Physics and Astronomy, University of Hawaii-Manoa, Honolulu, HI 96822, USA}
\date{}

\maketitle

\begin{abstract}
Coherent elastic neutrino-nucleus scattering consistent with the standard model has been observed by the COHERENT experiment.
We study nonstandard neutrino interactions using the detected spectrum. For the case in which the nonstandard interactions (NSI) are induced by a vector mediator lighter than 50~MeV, we obtain constraints on the coupling of the mediator. For a heavier mediator, we find that degeneracies between the NSI parameters severely weaken the constraints. However, these degeneracies do not affect COHERENT constraints on the effective NSI parameters for matter propagation in the Earth.
\end{abstract}

\newpage
\section{Introduction}
The COHERENT experiment has observed coherent elastic neutrino-nucleus scattering
 (CE$\nu$NS)~\cite{Akimov:2017ade} 43 years after its theoretical prediction in the standard model (SM)~\cite{Freedman:1973yd}.
For neutrino energies below a few tens of MeV, CE$\nu$NS occurs when the momentum transfer $Q$ is comparable to the inverse of the nuclear radius $R$, i.e., $QR\lesssim 1$. Compared to scattering of isolated nucleons, the cross section for coherent scattering off a nucleus is enhanced by the square of the number of neutrons in the nucleus. However, in spite of its large cross section, it is difficult to observe CE$\nu$NS because of the small momentum transfer 
involved.

The COHERENT experiment measures neutrinos from the Spallation Neutron Source (SNS) at Oak Ridge National Laboratory, using a sodium doped CsI scintillator that can detect nuclear recoil energies down to a few keV. They find 6.7$\sigma$ CL evidence for CE$\nu$NS in good
agreement with SM predictions, after fifteen months of data accumulation. 
The measurement of CE$\nu$NS not only completes the SM picture of neutrino interactions, but also provides a tool to study new physics beyond the SM, e.g., nonstandard neutrino interactions (NSI)~\cite{Barranco:2005yy, Scholberg:2005qs, Coloma:2017egw}, sterile neutrinos~\cite{Anderson:2012pn}, neutrino magnetic moment~\cite{Dodd:1991ni}, and light dark matter~\cite{deNiverville:2015mwa}.
 
The total number of events has been used in Refs.~\cite{Akimov:2017ade} and~\cite{Coloma:2017ncl} to constrain NSI under the contact interaction approximation. If the momentum transfer of CE$\nu$NS is comparable to the mediator mass, the shape of the spectrum is also modified. Consequently, constraints obtained using the contact approximation do not apply to the light mediator case. In this Letter, we use the spectrum of the CE$\nu$NS signal to constrain NSI including mediator effects. In Section~2, we describe our simulation of the COHERENT spectrum. In Section~3, we discuss the sensitivities to the nonstandard parameters for both the light and heavy mediators. We summarize our results in Section~4.

\section{COHERENT simulation}
The expected number of events for neutrino flavor $\alpha$ of recoil energy $E_r$ is
\begin{align}
\frac{dN_\alpha}{dE_r}=n_\text{N}\int dE_{\nu}\phi_\alpha(E_\nu)\frac{d\sigma_\alpha}{dE_r}(E_\nu)\,,
\end{align} 
where the total number of nucleons in the detector is $n_\text{N}=\frac{2m_\text{det}}{M_\text{CsI}}N_A$, with $m_\text{det}= 14.6$~kg being the detector mass, $M_\text{CsI}$ the molar mass of CsI, and $N_A$ the Avogadro constant.
Neutrinos at the SNS consist of a prompt component of monochromatic $\nu_\mu$ from the stopped pion decays, $\pi^+\to \mu^++\nu_\mu$, and two delayed components of $\bar{\nu}_\mu$ and $\nu_e$ from the subsequent muon decays, $\mu^+\to e^++\bar{\nu}_\mu+\nu_e$. The distribution of the total flux for each neutrino flavor is well-known and given by~\cite{Coloma:2017egw}
\begin{align}
\phi_{\nu_\mu}(E_\nu)&={\cal{N}}\delta\left(E_\nu-\frac{m_\pi^2-m_\mu^2}{2m_\pi}\right)\,,
\nonumber\\
\phi_{\bar{\nu}_\mu}(E_\nu)&={\cal{N}}\frac{64E_\nu^2}{m_\mu^3}\left(\frac{3}{4}-\frac{E_\nu}{m_\mu}\right)\,,
\nonumber\\
\phi_{\nu_e}(E_\nu)&={\cal{N}}\frac{192E_\nu^2}{m_\mu^3}\left(\frac{1}{2}-\frac{E_\nu}{m_\mu}\right)\,,
\end{align}
where the normalization factor is ${\cal{N}}=\frac{rN_\text{POT}}{4\pi L^2}$. Here $r=0.08$ is the number of neutrinos per flavor that are produced for each proton on target~\cite{Akimov:2017ade}. The total number of protons delivered to the mercury target is $N_\text{POT}=1.76\times 10^{23}$ and the distance between the source and the CsI detector is $L=19.3$~m~\cite{Akimov:2017ade}.

The differential cross section for a given neutrino flavor $\nu_\alpha$ in the SM is
\begin{align}
\frac{d\sigma_{\alpha }}{dE_r}=\frac{G_F^2}{2\pi}Q_{\alpha }^2F^2(2ME_r)M
\left(2-\frac{ME_r}{E_\nu^2}\right)\,,
\label{eq:CC}
\end{align}
where $M$ is the mass of the target nucleus, $F(Q^2)$ is the nuclear form factor, and the radiative corrections are neglected.
We take the nuclear form factor from Ref.~\cite{Klein:1999gv}. The effective charge in the SM is 
\begin{align}
Q_{\alpha,\text{SM} }^2=&\left(Zg_p^V+Ng_n^V\right)^2\,,
\end{align}
where $Z$ and $N$ are the number of protons and neutrons in the nucleus, and $g_p^V=\frac{1}{2}-2\sin^2\theta_W$ and $g_n^V=-\frac{1}{2}$ are the SM couplings of the $Z^0$ boson to the proton and neutron, with $\theta_W$ the weak mixing angle. 

\begin{figure}
\centering
\includegraphics[width=0.8\textwidth]{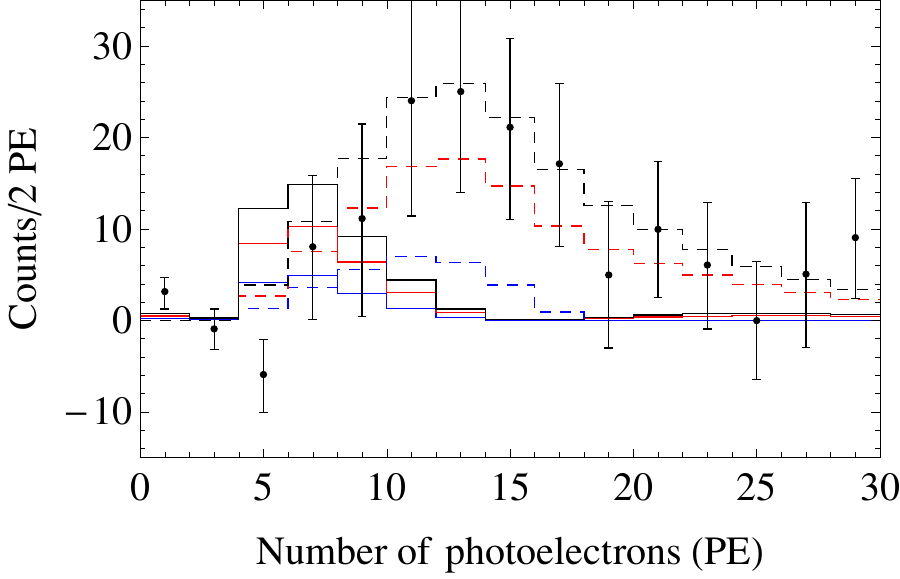}
\caption{The expected CE$\nu$NS events as a function of the number of photoelectrons. The dashed lines correspond to the SM, and the solid lines correspond to the NSI case with $M_{Z'}=10$ MeV and $g=10^{-4}$. The blue (red) [black] lines correspond to the $\nu_\mu$ ($\nu_\mu+\bar{\nu}_\mu$) [$\nu_\mu+\bar{\nu}_\mu+\nu_e$] contributions. }
\label{fig:spectra}
\end{figure}

We ignore the contribution to the cross section from the sodium dopant because of its extremely small fractional mass ($10^{-4}-10^{-5}$) in the CsI detector~\cite{Collar:2014lya}. Also, since the responses of Cs and I to a given neutrino flavor are almost identical due to very similar nuclear masses~\cite{Collar:2014lya}, we do not distinguish between Cs and I in our analysis. We adopt a simple relation between the observed number of photoelectrons (PE) and the nuclear recoil energy~\cite{Akimov:2017ade}:
\begin{align}
n_\text{PE}=1.17 \left(\frac{E_r}{ \text{keV}}\right)\,.
\end{align}   
After taking into account the surviving fraction of the CE$\nu$NS signals as a function of the number of photoelectrons given in Fig.~S9 in Ref.~\cite{Akimov:2017ade}, 
we show the expected CE$\nu$NS events as a function of the number of photoelectrons for the SM in Fig.~\ref{fig:spectra}. 

\section{Constraints on nonstandard neutrino interactions}
We now analyze the COHERENT spectrum to constrain vector NSI parameters that lead to an effective potential for neutrino propagation in matter. In principle, COHERENT data also constrain axial-vector NSI, but because large nuclei approximately conserve parity, the constraints are weak.
\subsection{Light mediator}
We first consider the case that NSI are induced by a light vector mediator $Z'$ with mass $M_{Z'}$ that is comparable to the square root of the momentum transfer. For simplicity, we assume that the $Z'$ has purely vector universal flavor-conserving couplings to neutrinos, first generation quarks and the muon. Then the effective charge in Eq.~(\ref{eq:CC}) can be written as
\begin{align}
Q_{\alpha,\text{NSI} }^2=\left[Z\bigg(g_p^V+\frac{3g^2}{2\sqrt{2}G_F(Q^2+M_{Z'}^2)}\bigg)
+N\bigg(g_n^V+\frac{3g^2}{2\sqrt{2}G_F(Q^2+M_{Z'}^2)}\bigg)\right]^2\,,
\label{eq:propagator}
\end{align}
where $Q^2=2ME_r$ is the square of the momentum transfer. 

To evaluate the statistical significance, we define
\begin{align}
\chi^2=\sum_i\left[\frac{N_\text{exp}^i-N_\text{NSI}^i(1+\alpha)}{\sigma_{\text{stat}}^{i}}\right]^2+\left(\frac{\alpha}{\sigma_\alpha}\right)^2\,,
\end{align}
where $N_\text{exp}^i$ ($N_\text{NSI}^i$) is the number of observed (predicted) events per bin, $\sigma_{\text{stat}}^{i}$ is the statistical uncertainty, and the total normalization uncertainty is $\sigma_\alpha=0.28$, which incorporates the neutrino flux, form factor, quenching factor and signal acceptance uncertainties~\cite{Akimov:2017ade}. We extract $N_\text{exp}^i$ and $\sigma_{\text{stat}}^{i}$ from the top right panel of Fig. 3 in Ref.~\cite{Akimov:2017ade}, and consider 12 bins in the $6\leq \text{PE}<30$ range, and ignore the small background from prompt neutrons. 

\begin{figure}
\centering
\includegraphics[width=0.6\textwidth]{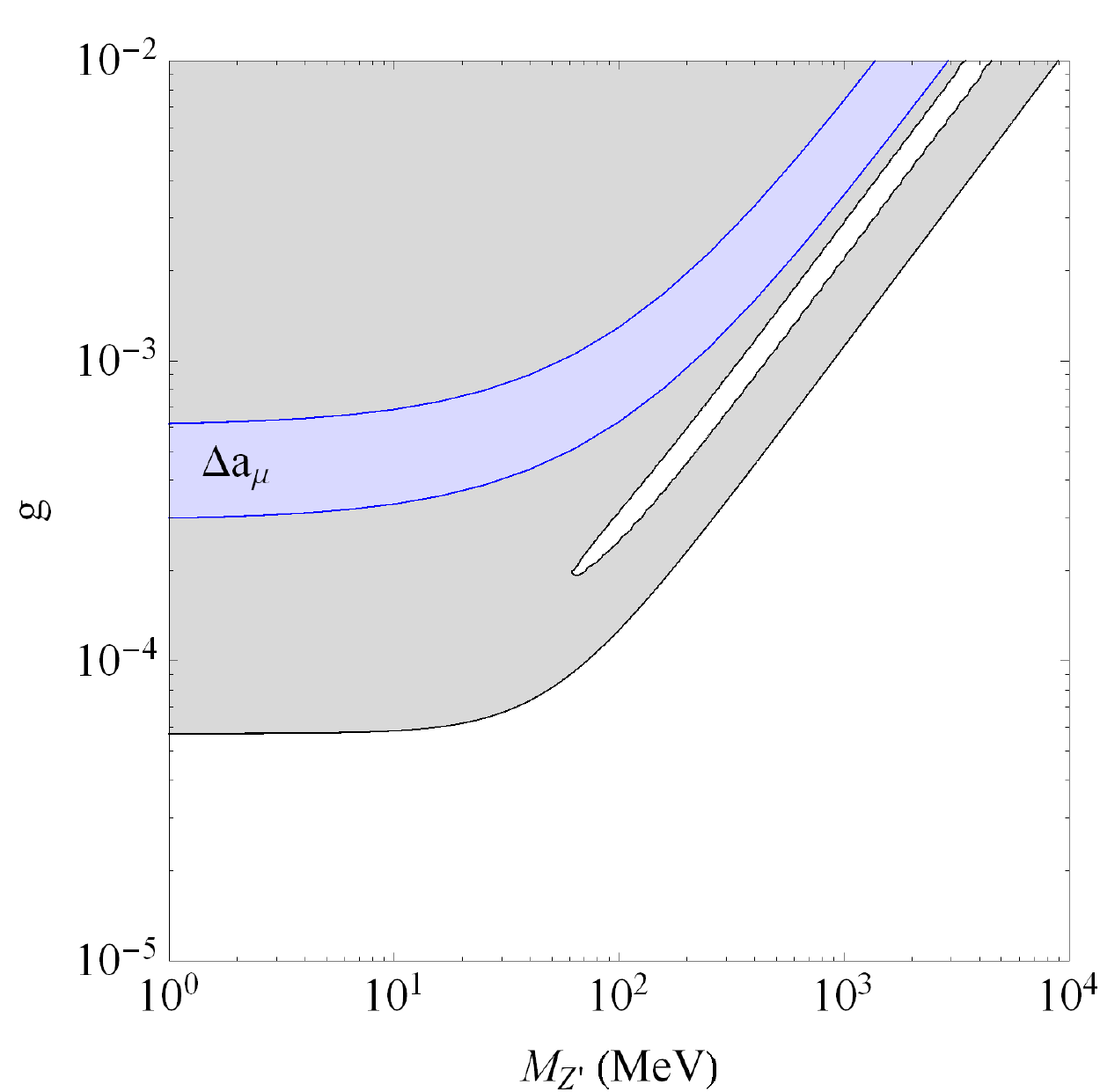}
\caption{The $2\sigma$ exclusion region in the $(M_{Z'}, g)$ plane from the COHERENT data. The $2\sigma$ allowed region that explains the discrepancy in the anomalous magnetic moment of the muon ($\Delta a_\mu=(29\pm 9)\times 10^{-10}$~\cite{jeg}) is shown for comparison.}
\label{fig:lm}
\end{figure}

We scan over possible values of the coupling $g$ and the mediator mass $M_{Z'}$, and show the $2\sigma$ limits in the $(M_{Z'}, g)$ plane in Fig.~\ref{fig:lm}. The $2\sigma$ allowed region that explains the discrepancy in the anomalous magnetic moment of the muon~\cite{jeg} is also shown for comparison. We see that a light mediator that can explain the discrepancy in the anomalous magnetic moment of the muon is disfavored.

The shape of the limit curve in Fig.~\ref{fig:lm} can be understood from the propagator in Eq.~(\ref{eq:propagator}), in which the NSI contribution is proportional to $\frac{g^2}{2ME_r+M_{Z'}^2}$. For a very light mediator, i.e., $M_{Z'}\ll \sqrt{2ME_r}\sim 50$~MeV, the limit is only sensitive to the coupling $g$. Note that since the momentum transfer in coherent forward scattering is zero, the NSI matter effect for neutrino propagation is sensitive to $\frac{g^2}{M_{Z'}^2}$~\cite{Farzan:2015hkd}, and the constraint does not apply to matter NSI induced by a very light mediator. For a heavy mediator, i.e., $M_{Z'}\gg \sqrt{2ME_r}$,  NSI do not change the shape of the spectra, and the limit is dependent on the ratio $\frac{g}{M_{Z'}}$. There is also a degenerate region that is not excluded by current data.
Since the data are consistent with the SM, the degenerate region can be understood by the relation, $Q_{\alpha,\text{NSI}}=-Q_{\alpha,\text{SM}}$, i.e.,
\begin{align}
\frac{g^2}{M_{Z'}^2}=-\frac{4\sqrt{2}(Zg_p^V+Ng_n^V)}{3(Z+N)}G_F \,,
\end{align} 
which holds for all $E_r$ bins when $M_{Z'}\gg \sqrt{2ME_r}$. For a light mediator, the spectral shapes are modified by NSI (see the solid lines in Fig.~\ref{fig:spectra} for example), which breaks the degeneracy.

\subsection{Heavy mediator}
For a heavy mediator, matter NSI can be described by four-fermion contact operators of the form~\cite{Wolfenstein:1977ue}
\bea
  \label{eq:NSI}
  \mathcal{L}_\text{NSI} = - \sqrt{2}G_F
   \epsilon^{fV}_{\alpha\beta} 
        \left[ \overline{\nu}_{\alpha L} \gamma^{\rho} \nu_{\beta L} \right] 
        \left[ \bar{f} \gamma_{\rho}f \right]\,,
\eea
where $\alpha, \beta=e, \mu, \tau$, $f=u,d$, and the strength of the new interaction
$\epsilon^{fV}_{\alpha\beta}$ is parameterized in units of $G_F$. As before, we consider NSI couplings to first generation quarks but not to electrons. We take the phases of the off-diagonal NSI parameters to be 0. For a heavy mediator, the effective charge in Eq.~(\ref{eq:CC}) is
\begin{align}
Q_{\alpha }^2=\left[Z(g_p^V+2\epsilon_{\alpha\alpha}^{uV}+\epsilon_{\alpha\alpha}^{dV})
+N(g_n^V+\epsilon_{\alpha\alpha}^{uV}+2\epsilon_{\alpha\alpha}^{dV})\right]^2
+\sum_{\beta\neq\alpha}\left[Z(2\epsilon_{\alpha\beta}^{uV}+\epsilon_{\alpha\beta}^{dV})
+N(\epsilon_{\alpha\beta}^{uV}+2\epsilon_{\alpha\beta}^{dV})\right]^2\,.
\end{align}
\begin{figure}
\centering
\includegraphics[width=0.4\textwidth]{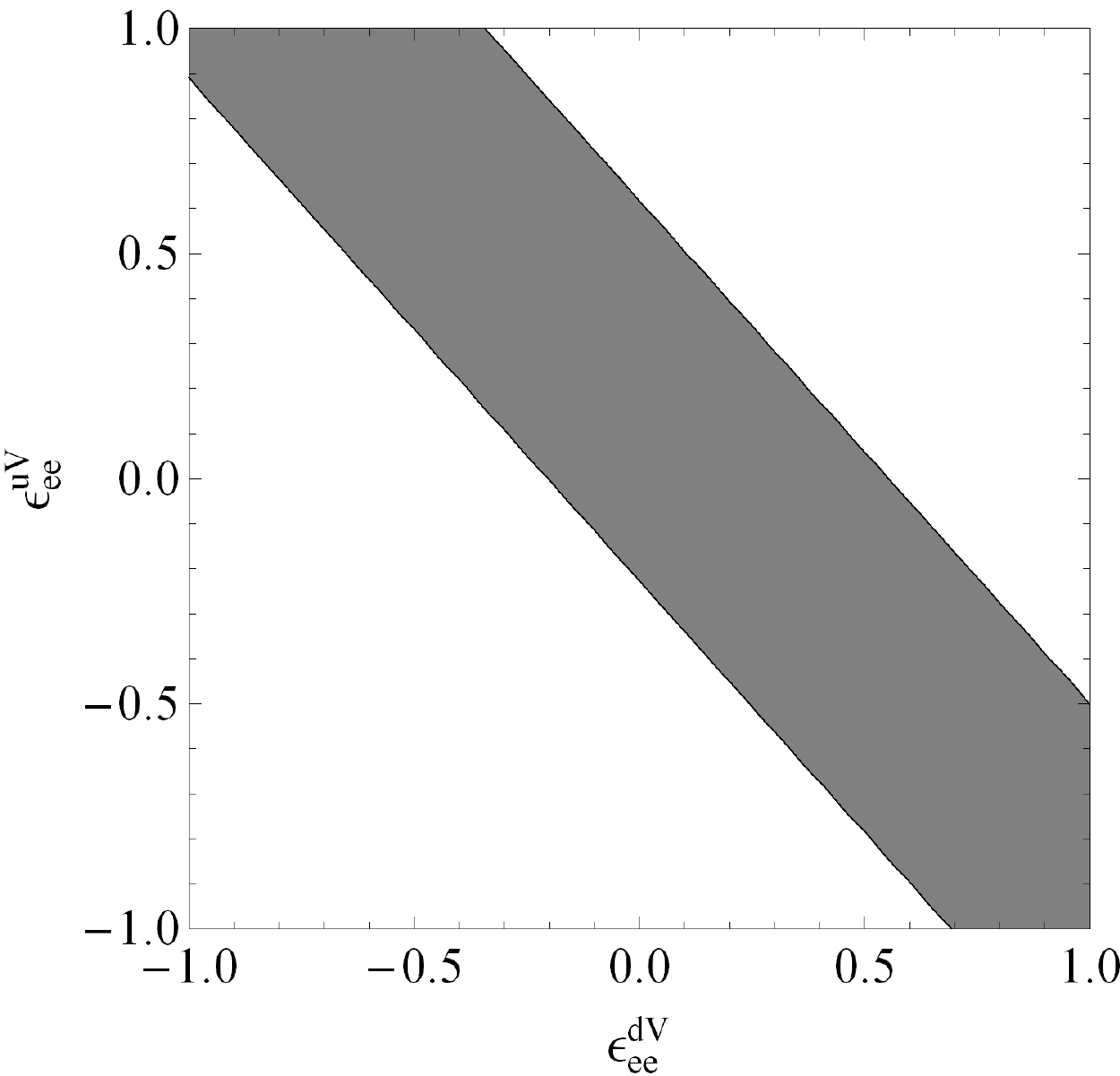}
\includegraphics[width=0.4\textwidth]{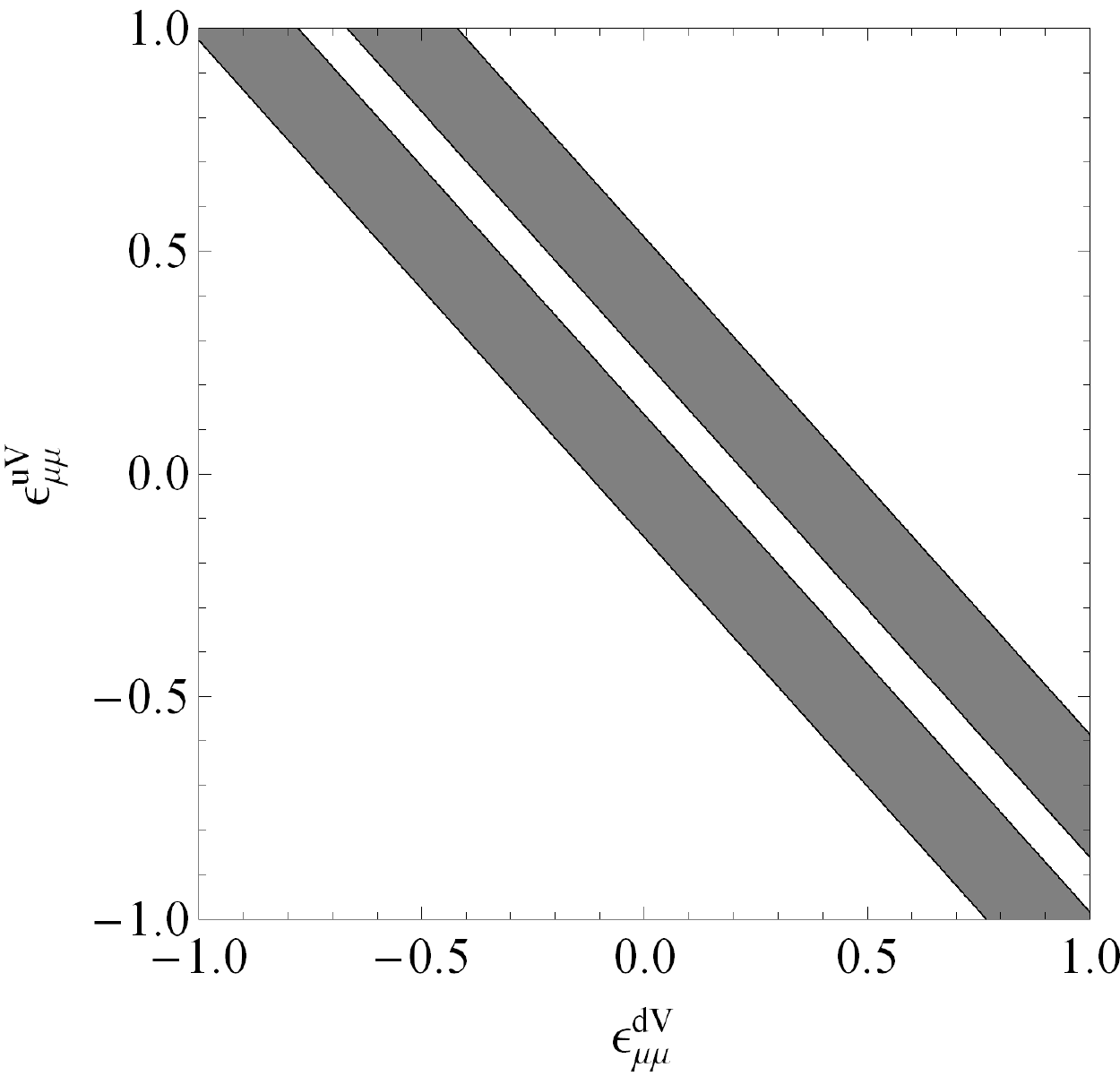}
\includegraphics[width=0.4\textwidth]{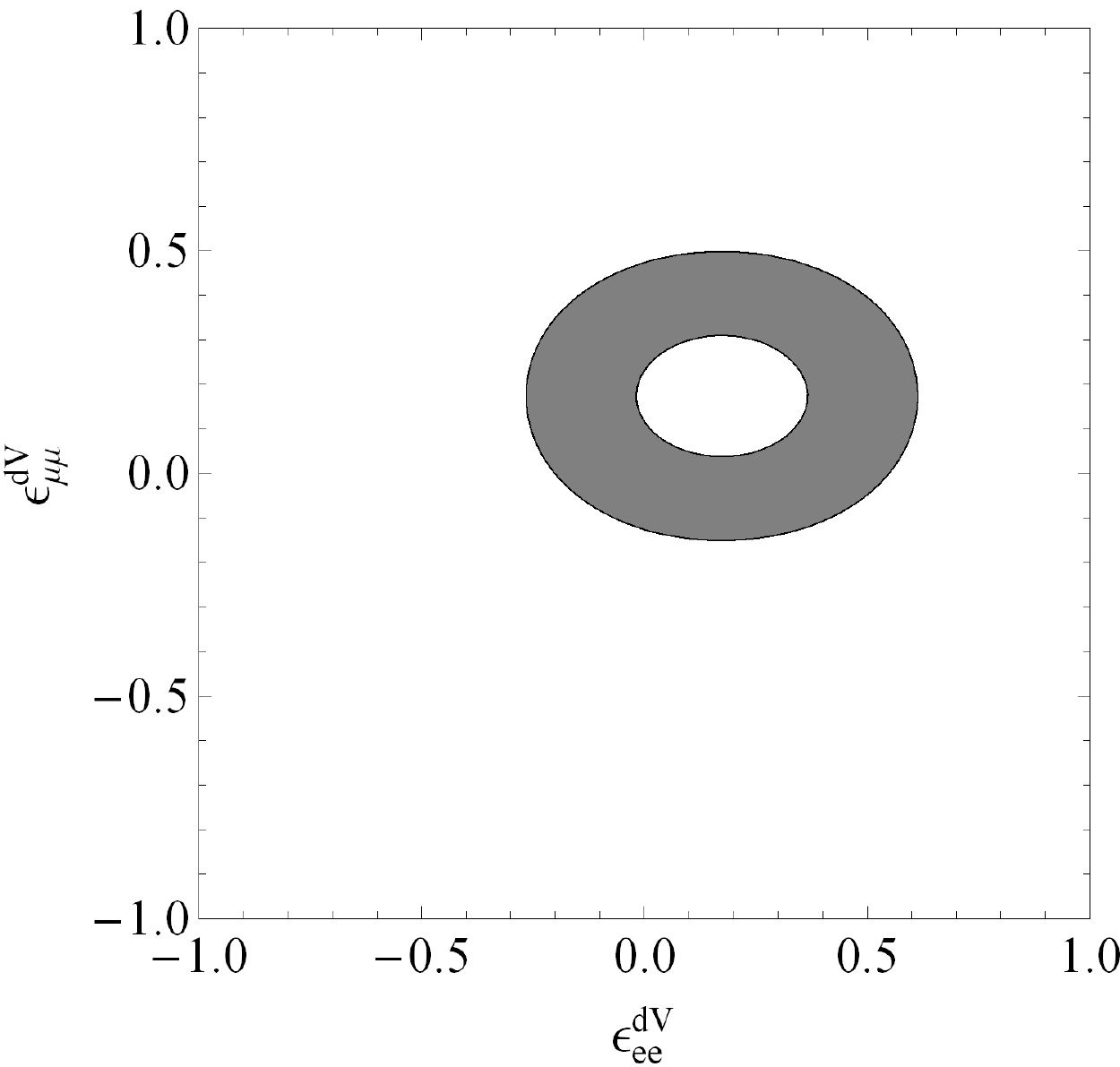}
\includegraphics[width=0.4\textwidth]{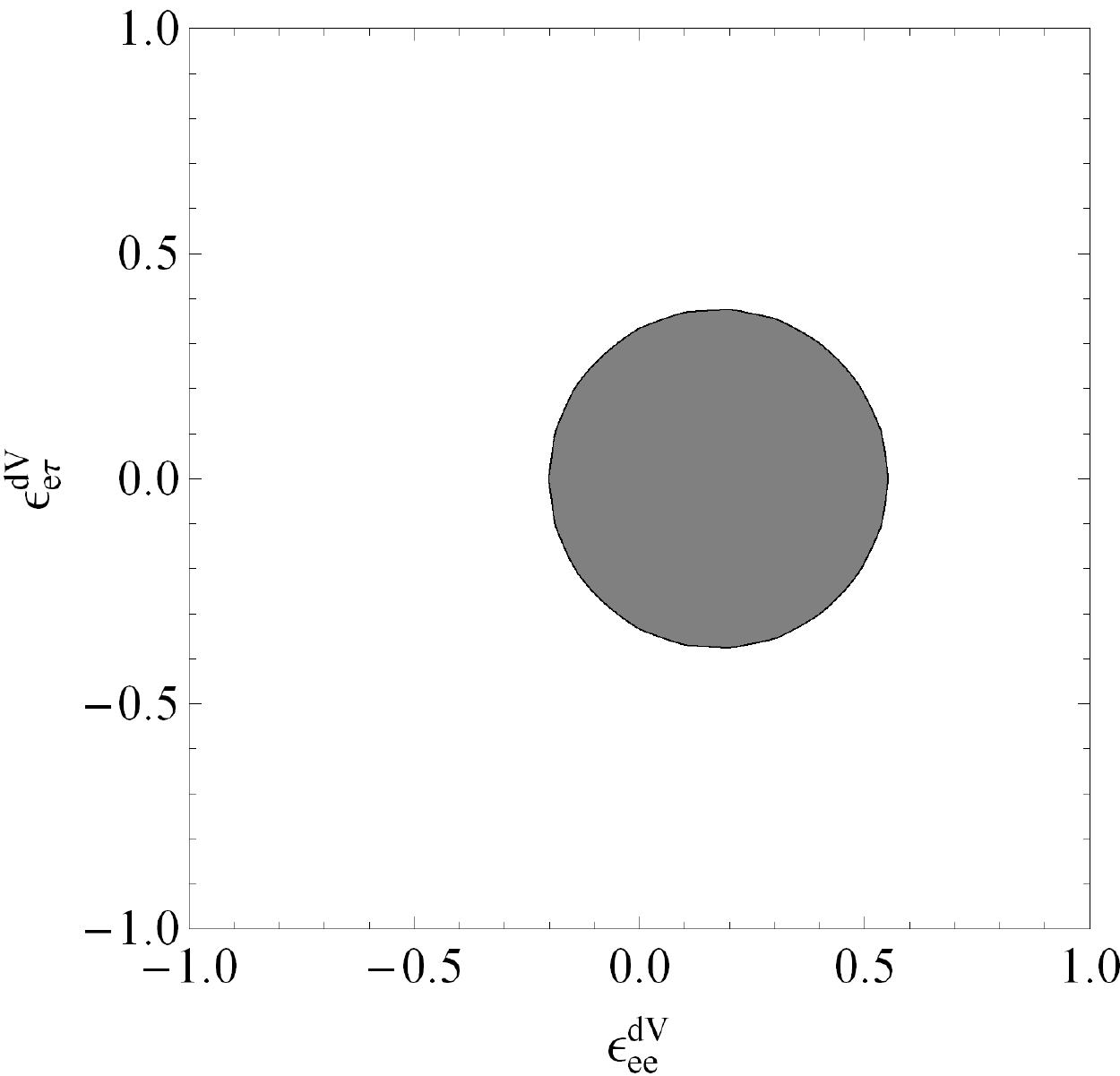}
\caption{The $90\%$ CL regions in the NSI parameter space allowed by COHERENT data. The NSI parameters not shown in each graph are assumed to be zero. }
\label{fig:nsi}
\end{figure}
We consider four cases with only two nonzero NSI parameters for simplicity:
\begin{itemize}
\item[(a)] $\epsilon_{ee}^{uV}\neq 0$, $\epsilon_{ee}^{dV}\neq 0$. 
In this case, only the electron neutrinos are affected, and the $90\%$~CL allowed region in the ($\epsilon_{ee}^{dV}$, $\epsilon_{ee}^{uV}$) plane is shown in the top-left panel of Fig.~\ref{fig:nsi}. Since the data are consistent with the SM, the allowed regions can be understood by the relation, $Zg_p^V+Ng_n^V=\pm \left[Z(g_p^V+2\epsilon_{ee}^{uV}+\epsilon_{ee}^{dV})
+N(g_n^V+\epsilon_{ee}^{uV}+2\epsilon_{ee}^{dV})\right]$, which yields two linear bands in the ($\epsilon_{ee}^{dV}$, $\epsilon_{ee}^{uV}$) parameter space~\cite{Scholberg:2005qs}. Because the $\nu_e$ contribution to the total neutrino flux is small, the two bands merge into a single band. In principle, multiple detector elements should break the degeneracy, but since the Cs and I nuclei have very similar nucleon masses, the degeneracy is unbroken.  
\item[(b)] $\epsilon_{\mu\mu}^{uV}\neq 0$, $\epsilon_{\mu\mu}^{dV}\neq 0$. 
The $90\%$ CL allowed regions in this case are shown in the top-right panel of Fig.~\ref{fig:nsi}.
This case is similar to the electron neutrino case, except that both $\nu_\mu$ and $\bar{\nu}_\mu$ are affected. Since the $\nu_\mu+\bar{\nu}_\mu$ flux contribution is more than twice that of $\nu_e$, the two bands do not overlap. 
\item[(c)] $\epsilon_{ee}^{dV}\neq 0$, $\epsilon_{\mu\mu}^{dV}\neq 0$. 
The $90\%$ CL allowed region in this case is shown in the bottom-left panel of Fig.~\ref{fig:nsi}. All three neutrino components are affected. The results for $\epsilon_{ee}^{uV}= 0$ and for  $\epsilon_{\mu\mu}^{dV}=0$ are consistent with those in case (a) and case (b), respectively.
\item[(d)] $\epsilon_{ee}^{dV}\neq 0$, $\epsilon_{e\tau}^{dV}\neq 0$. 
In this case, only the electron neutrinos are affected, and the $90\%$~CL allowed region is shown in the bottom-right panel of Fig.~\ref{fig:nsi}. The expected allowed region is given by the relation, $\left(Zg_p^V+Ng_n^V\right)^2= \left[Z(g_p^V+\epsilon_{ee}^{dV})+N(g_n^V+2\epsilon_{ee}^{dV})\right]^2+\left[Z\epsilon_{e\tau}^{dV}
+2N\epsilon_{e\tau}^{dV}\right]^2$, which yields a region between two ellipses (an annulus)~\cite{Scholberg:2005qs}. We see a single ellipse due to the small $\nu_e$ flux. 
\end{itemize}

\begin{figure}
\centering
\includegraphics[width=0.5\textwidth]{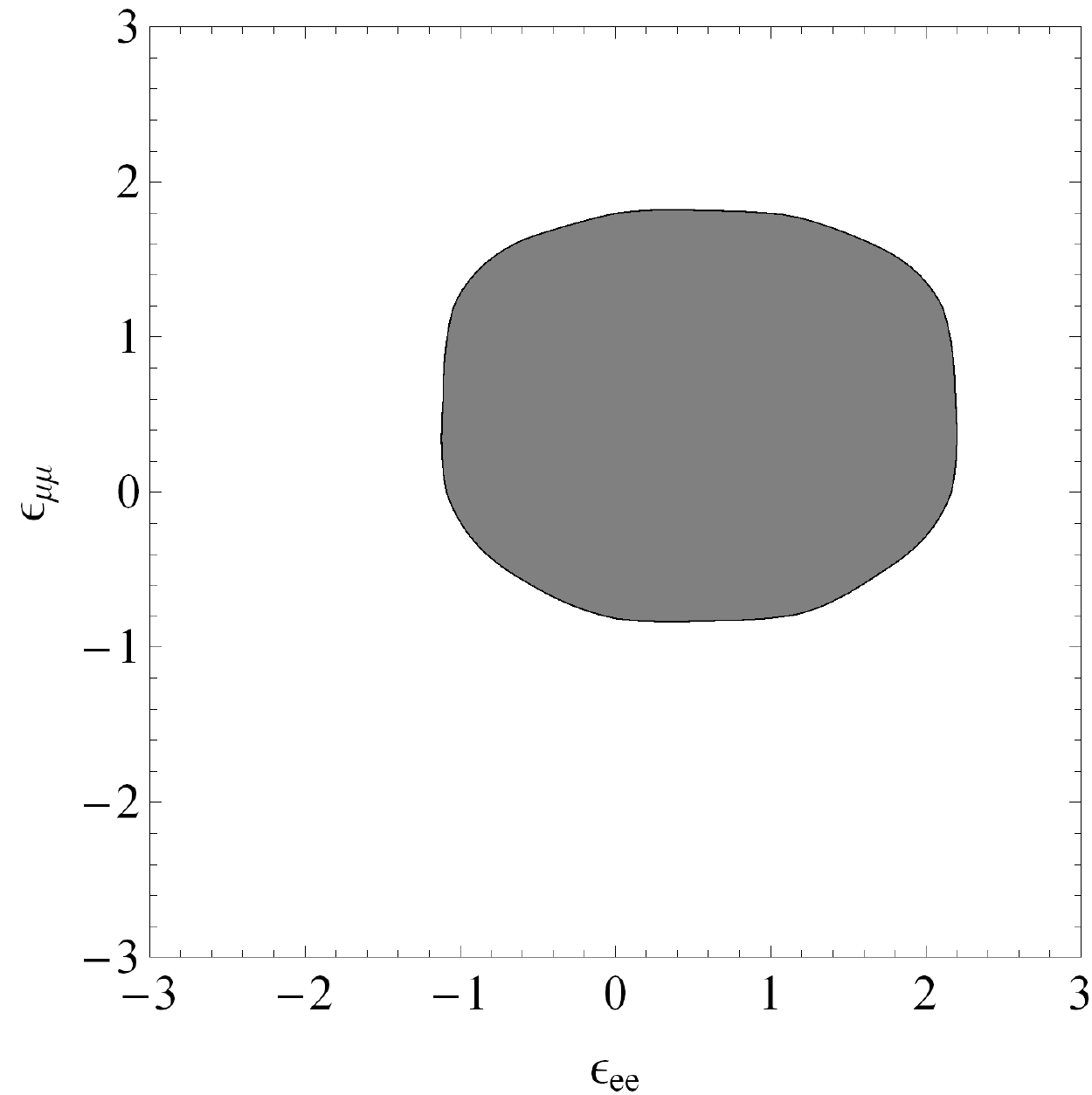}
\caption{The $90\%$ allowed regions in the ($\epsilon_{ee}$, $\epsilon_{\mu\mu}$) plane from the COHERENT data. }
\label{fig:eemm}
\end{figure}

Degeneracies between different combinations of NSI parameters, especially the cancellation between the NSI coupling to up and down quarks, permit large values of NSI parameters. However, the effective NSI paremeters in Earth matter are dependent on the sum of the up-type and down-type NSI parameters, i.e.,~\cite{rk}
\begin{align}
\epsilon_{\alpha\alpha}\approx 3(\epsilon_{\alpha\alpha}^{uV}+\epsilon_{\alpha\alpha}^{dV})\,.
\end{align}
Thus, COHERENT constraints on the effective NSI parameters do not depend on the cancellation between the up-type and down-type NSI parameters. As an illustration, we scan over all possible values of $\epsilon_{ee}^{dV}$, $\epsilon_{ee}^{uV}$, $\epsilon_{\mu\mu}^{dV}$, $\epsilon_{\mu\mu}^{uV}$, and show the projected $90\%$ CL allowed regions in the ($\epsilon_{ee}$, $\epsilon_{\mu\mu}$) plane in Fig.~\ref{fig:eemm}. At $90\%$ CL, the  effective NSI parameters lie in the ranges, 
\begin{equation}
-0.95 \leq \epsilon_{ee} \leq 1.95\,,\ \ \ \ \ \ -0.66 \leq \epsilon_{\mu\mu} \leq 1.57\,.
\end{equation}
%
%

\section{Summary}
We analyzed the spectrum of coherent elastic neutrino-nucleus scattering observed by the COHERENT experiment to constrain nonstandard neutrino interactions. For NSI induced by a vector mediator lighter than 50~MeV, COHERENT data only constrain the mediator coupling $g$. Since the NSI matter effect in neutrino propagation depends on $\frac{g^2}{M_{Z'}^2}$, the constraint does not apply to matter NSI induced by a very light mediator. For a heavier mediator, the COHERENT constraints are weakened by degeneracies between different combinations of NSI parameters. In particular, a cancellation between the NSI couplings to up and down quarks allows very large NSI parameters. However, COHERENT data place meaningful constraints on the effective NSI parameters in Earth matter since they depend on the sum of the up-type and down-type NSI parameters.

\vspace{0.1 in}
{\it Acknowledgments.} We thank K.~Scholberg for helpful correspondence. This research was supported in part by the
U.S. DOE under Grant No. DE-SC0010504.


\end{document}